\documentclass[sigconf]{acmart}

\usepackage{subcaption}
\usepackage[subtle]{savetrees}
\usepackage{enumitem}
\AtBeginDocument{%
  \providecommand\BibTeX{{%
    \normalfont B\kern-0.5em{\scshape i\kern-0.25em b}\kern-0.8em\TeX}}}

\copyrightyear{2021}

\begin{document}

\title{Conduit: A C++ Library for Best-effort High Performance Computing}

\author{Matthew Andres Moreno}
\email{mmore500@msu.edu}
\affiliation{%
  \institution{Michigan State University}
  \city{East Lansing}
  \state{Michigan}
  \country{United States}
  \postcode{48824}
}

\author{Santiago Rodriguez Papa}
\email{rodri816@msu.edu}
\affiliation{%
  \institution{Michigan State University}
  \city{East Lansing}
  \state{Michigan}
  \country{United States}
  \postcode{48824}
}

\author{Charles Ofria}
\email{ofria@msu.edu}
\affiliation{%
  \institution{Michigan State University}
  \city{East Lansing}
  \state{Michigan}
  \country{United States}
  \postcode{48824}
}

\renewcommand{\shortauthors}{Moreno et al.}

\begin{abstract}


Developing software to effectively take advantage of growth in parallel and distributed processing capacity poses significant challenges.
Traditional programming techniques allow a user to assume that execution, message passing, and memory are always kept synchronized.
However, maintaining this consistency becomes increasingly costly at scale. 
One proposed strategy is ``best-effort computing'', which relaxes synchronization and hardware reliability requirements, accepting nondeterminism in exchange for efficiency.
Although many programming languages and frameworks aim to facilitate software development for high performance applications, existing tools do not directly provide a prepackaged best-effort interface.
The Conduit C++ Library aims to provide such an interface for convenient implementation of software that uses best-effort inter-thread and inter-process communication.
Here, we describe the motivation, objectives, design, and implementation of the library.
Benchmarks on a communication-intensive graph coloring problem and a compute-intensive digital evolution simulation show that Conduit's best-effort model can improve scaling efficiency and solution quality, particularly in a distributed, multi-node context.

\end{abstract}

%

\maketitle

\section{Introduction}

The parallel and distributed processing capacity of high-performance computing clusters continues to grow rapidly and enable profound scientific and industrial innovations \cite{gagliardi2019international}.    
Hardware advances afford great opportunity, but also pose a serious challenge: developing approaches to effectively harness it.
As HPC systems scale, deterministic algorithms depending on global synchronization become increasingly costly \cite{gropp2013programming,dongarra2014applied}.

Best-effort computing, where data dependencies are relaxed to reduce synchronization \cite{chakradhar2010best}, can improve scalability \cite{meng2009best}.
Evolutionary algorithms typically perform perform a search or optimization with many acceptable results using pseudo-stochastic methods.
Algorithms with these properties can often tolerate significant perturbations, including relaxation of synchronization requirements in the underlying algorithm.
For example, island model genetic algorithms have been shown to perform well with asynchronous migration \cite{izzo2009parallel}.
While not the focus of this paper, the original interest that motivated development of the Conduit library stems from work exploring open-ended evolution.
Researchers in this domain study long-term dynamics of evolutionary systems in order to understand factors that affect these systems' potential to generate ongoing novelty \cite{taylor2016open}.
Some recent evidence suggests that the generative potential of systems devised to study open-ended evolution is --- at least in part --- meaningfully constrained by available compute resources \cite{channon2019maximum}.
Such observations raise the question of how to design parallel and distributed open-ended evolution systems.

The concept of indefinite scalability was developed to describe constraints distributed systems would face at the asymptote of technological (and even physical) constraints.
Indefinite scalability theory posits that such systems would necessitate
\begin{itemize}
    \item asynchronous operation,
    \item decentralized networking,
    \item interchangeable components (i.e., no global identifiers),
    \item and graceful degradation under hardware failure \cite{ackley2011pursue}.
\end{itemize}
Although bespoke experimental hardware such as Illuminato X machina has been developed to demonstrate aspects of the indefinite scalability paradigm \cite{ackley2011homeostatic}, practical contemporary work scaling up evolving systems necessitates working with commercially-available hardware.
However, existing distributed computing frameworks for contemporary high-performance computing hardware do not explicitly expose a convenient best-effort interface.
In our current work, we focus on asynchronous operation and decentralization.
We leave system robustness and resistance to hardware degradation to future work.

The Message Passing Interface (MPI) standard \cite{gropp1996high} represents the mainstay for high-performance computing applications.
This standard exposes communication primitives directly to the end user.
MPI's nonblocking communication primitives, in particular, are sufficient to program distributed computations with relaxed synchronization requirements.
Although its explicit, imperative nature enables precise control over execution, it also poses significant expense in terms of programability.
This cost manifests in terms of programmer productivity, domain knowledge requirements, software quality, and difficulty tuning for performance due to program brittleness \cite{gu2019comparative, tang2014mpi}.

In response to programmability concerns, many frameworks have arisen to offer useful parallel and distributed programming abstractions.
Task-based frameworks such as Charm++ \cite{kale1993charm++}, Legion \cite{bauer2012legion}, Cilk \cite{blumofe1996cilk}, and Threading Building Blocks (TBB) \cite{reinders2007intel} describe the dependency relationships among computational tasks and associated data and relies on an associated runtime to automatically schedule and manage execution.
These frameworks assume a deterministic relationship between tasks.
In a similar vein, programming languages and extensions like Unified Parallel C (UPC) \cite{el2006upc} and Chapel \cite{chamberlain2007parallel} rely on programmers to direct execution, but equips them with powerful abstractions, such as global shared memory.
However, Chapel's memory model explicitly forbids data races and UPC ultimately relies on a barrier model for data transfer.

\section{Library Design}

   \begin{figure}[thpb]
      \centering
      \includegraphics[width=0.8\linewidth]{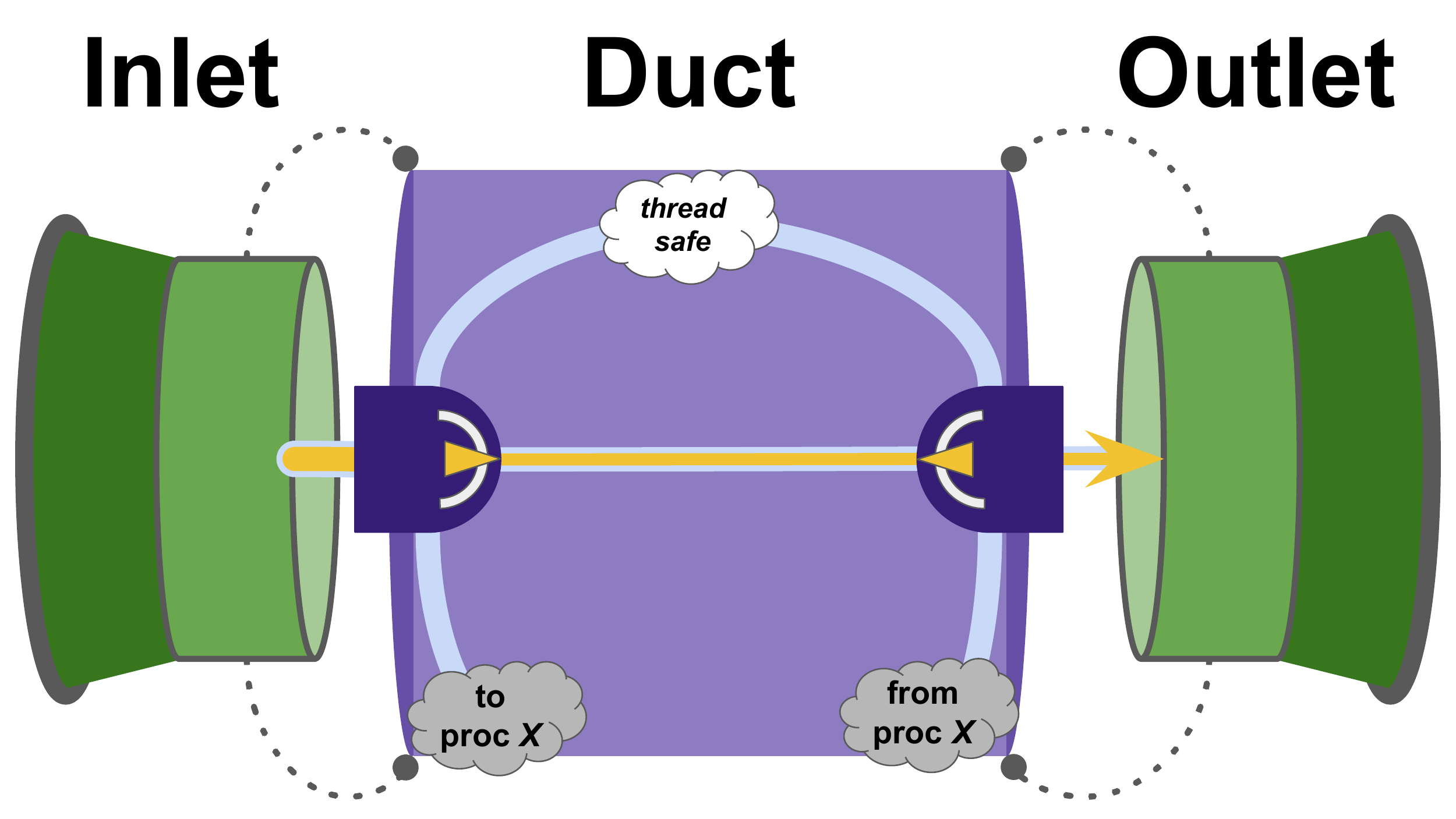}
      \caption{Schematic of Conduit's \texttt{Inlet} and \texttt{Outlet} object scheme.
      Intra-thread, inter-thread, or inter-process communication behavior is performed by an underlying \texttt{Duct} object.
      The underlying communication mechanism used is transparent to the end-user.
      }
      \label{fig:conduit}
   \end{figure}

The Conduit C++ Library aims to compliment the parallel and distributed programming ecosystem by providing an abstracted best-effort interface to application programmers.
Under Conduit's interface, more recent messages may preempt existing ones, messages may be dropped under backlog conditions, and read operations may opt to view the most recently received message in lieu of waiting for an expected message.

Conduit represents communication in terms of a paired \texttt{Inlet}, which accepts messages, and \texttt{Outlet}, which dispenses messages.
An \texttt{Inlet} and \texttt{Outlet} may exchange messages via an intra-thread, inter-thread, or inter-process communication procedure, depending on the runtime state of an underlying \texttt{Duct} object.
Figure \ref{fig:conduit} provides a schematic overview.
The implementation of intra-thread, inter-thread, and inter-process communication procedures may be configured at compile-time.
Conduit provides a library of intra-thread, inter-thread, and inter-process implementations to choose from and allows end-users to build their own.

The \texttt{Inlet} provides a non-blocking \texttt{TryPut()} method, which attempts to queue a message for its corresponding \texttt{Outlet} but may drop it under backlog conditions, as well as a \texttt{Put()} method, which block under backlog conditions until buffer space is available to queue the message is available.
The \texttt{Outlet} provides a \texttt{TryStep()} and \texttt{Jump()} methods to load the next or latest message, respectively.
If no new messages are available, the last-received message will be accessed.
The \texttt{Outlet} also provides a \texttt{Step()} method, which will block until a new message is received.
At run time, \texttt{Duct}s can be created or modified to perform intrathread, interthread, or interprocess communication. 

In addition to this granular connection-level interface, Conduit provides a network-level interface where the user defines their computation in terms of a directed graph.
In this network topology, nodes represent simulation elements and edges represent communication channels.
The library will assign nodes in that topology to available threads across available processes and automatically instantiate appropriate conduits.
Individual threads of execution can then launch and freely process computational updates on their assigned nodes while receiving messages from nodes assigned to other processes as they become available.
Conduit provides several pre-defined standard topologies and node-assignment algorithms.
However, users can also opt to use the NetworkX graph library \cite{hagberg2008exploring} to generate arbitrary topologies and the METIS software package \cite{gupta1997fast} to automatically balance expected load across available threads and processes while minimizing inter-process and inter-thread communication.

Defining program logic in terms of atomic, inter-communicating simulation elements provides significant programmability advantages.
Such code can be written in terms of interactions between intuitive domain-specific objects (e.g., digital organisms that interact with one another, subpopulations with migration, etc.), while leaving the task of mapping onto hardware resources to automatic management by the underlying framework.
However, a naive implementation of this approach would entail significant inefficiency, particularly with inter-process communication which has a large overhead.
Imagine, for example, a processes dispatching independent MPI calls for each communication channel between another process and the thousands of atomic simulation elements it holds.
Conduit addresses this issue by providing duct implementations that automatically consolidate messages between processes into single MPI calls.
These implementations support a ``pooling'' mechanism, in which a consolidated MPI call is dispatched once each constituent \texttt{Inlet} has received a single message, and a ``aggregation'' mechanism, in which arbitrary numbers of messages can be contributed from each \texttt{Inlet}.

Conduit is made freely available under a MIT License as a header-only C++17 software package at \url{https://github.com/mmore500/conduit}.
Conduit was built using the cereal C++11 Library for serialization \cite{voorhies2017cereal} and the Empirical C++ Library \cite{charles_ofria_2019_2575607}.

\section{Methods}

We performed two benchmarks to compare the performance of Conduit's best-effort approach to a traditional synchronous model.
We tested our benchmarks across both a multithread, shared-memory context and a distributed, multinode context.
In each hardware context, we assessed performance on two algorithmic contexts: a communication-intensive distributed graph coloring problem (Section \ref{sec:graph_coloring_benchmark}) and a compute-intensive digital evolution simulation (Section \ref{sec:digital_evolution_benchmark}).
The latter benchmark --- presented in Section \ref{sec:digital_evolution_benchmark} --- grew out of the original work developing the Conduit library to support large-scale experimental systems to study open-ended evolution.
The former benchmark --- presented in Section \ref{sec:graph_coloring_benchmark} --- complements the first by providing a clear definition of solution quality.
Metrics to define solution quality in the open-ended digital evolution context remain a topic of active research.

\subsection{ Graph Coloring Benchmark } \label{sec:graph_coloring_benchmark}

The graph coloring benchmark employs a graph coloring algorithm designed for distributed WLAN channel selection \cite{leith2012wlan}.
In this algorithm, nodes begin by randomly choosing a color.
Each computational update, nodes test for any neighbor with the same color.
If and only if a conflicting neighbor is detected, nodes randomly select another color.
The probability of selecting each possible color is stored in array associated with each node.
Before selecting a new color, the stored probability of selecting the current (conflicting) color is decreased by a multiplicative factor $b$.
We used $b=0.1$, as suggested by Leith et al.
Likewise, the stored probability of selecting all others is increased by a multiplicative factor.
Regardless of whether their color changed, nodes transmit their current color to their neighbor.

Our benchmarks focus on weak scalability, using a fixed problem size of 2048 graph nodes per thread or process.
These nodes were arranged in a two-dimensional grid topology where each node had three possible colors and four neighbors.
We implement the algorithm with a single Conduit communication layer carrying graph color as an unsigned integer.
We used Conduit's built-in pooling feature to consolidate color information into a single MPI message between pairs of communicating processes each update. 
We performed five replicates, each with a five second simulation runtime.
Solution error was measured as the number of graph color conflicts remaining at the end of the benchmark. 

\subsection{ Digital Evolution Benchmark } \label{sec:digital_evolution_benchmark}

The digital evolution benchmark runs the DISHTINY (DIStributed Hierarchical Transitions in Individuality) artificial life framework.
This system is designed to study major transitions in evolution, events where lower-level organisms unite to form a self-replicating entity.
The evolution of multicellularity and eusociality exemplify such transitions. 
Previous work with DISHTINY has explored methods for selecting traits characteristic of multicellularity such as reproductive division of labor, resource sharing within kin groups, resource investment in offspring, and adaptive apoptosis \cite{moreno2019toward}.

DISHTINY simulates a fixed-size toroidal grid populated by digital cells.
Cells can sense attributes of their immediate neighbors, can communicate with those neighbors through arbitrary message passing, and can interact with neighboring cells cooperatively through resource sharing or competitively through antagonistic competition to spawn daughter cells into limited space.
This cell behavior is controlled by SignalGP event-driven linear genetic programs \cite{lalejini2018evolving}.
Full details of the DISHTINY simulation are available in \cite{moreno2021exploring}.

We use Conduit-based messaging channels to manage all interactions between neighboring cells.
During a computational update, each cell advances its internal state and pushes information about its current state to neighbor cells.
Several independent messaging layers handle disparate aspects of cell-cell interaction, including
\begin{itemize}
  \item Cell spawn messages, which contain arbitrary-length genomes (seeded at 100 12-byte instructions with a hard cap of 1000 instructions). These are handled every 16 updates and use Conduit's built-in aggregation support for inter-process transfer.
  \item Resource transfer messages, consisting of a 4-byte float value. These are handled every update and use Conduit's built-in pooling support for inter-process transfer.
  \item Cell-cell communication messages, consisting of arbitrarily many 20-byte packets dispatched by genetic program execution. These are handled every 16 updates and use Conduit's built-in aggregation support for inter-process transfer.
  \item Environmental state messages, consisting of a 216-byte struct of data. These are handled every 8 updates and use conduit's built-in pooling support for inter-process transfer.
  \item Multicellular kin-group size detection messages, consisting of a 16-byte bitstring. These are handled every update and use Conduit's built-in pooling support for inter-process transfer.
\end{itemize}

Implementing all cell-cell interaction via Conduit-based messaging channels allows the simulation to be parallelized down to the granularity, potentially, of individual cells.
However, in practice, for this benchmarking we assign 3600 cells to each thread or process.
Because all cell-cell interactions occur via Conduit-based messaging channels, logically-neighboring cells can interact fully whether or not they are located on the same thread or process (albeit with potential irregularities due to best-effort limitations). 
An alternate approach to evolving large populations might be an island model, where Conduit-based messaging channels would be used solely to exchange genomes between otherwise independent populations \cite{bennett1999building}.
However, we chose to instead parallelize DISHTINY as a unified spatial realm in order to enable parent-offspring interaction and leave the door open for future work with multicells that exceed the scope of an individual thread or process.

\subsection{Asynchronicity Modes} \label{sec:asynchronicity_modes}

\begin{table}
  \begin{tabular}{ccl}
    \toprule
    Mode & Description\\
    \midrule
    0 & Barrier sync every update\\
    1 & Rolling barrier sync\\
    2 & Fixed barrier sync\\
    3 & No barrier sync\\
    4 & No inter-cpu communication\\
  \bottomrule
\end{tabular}
  \caption{Asynchronicity modes used for benchmarking experiments, arranged from most to least synchronized.}
  \label{tab:asynchronicity_modes}
\end{table}

For both benchmarks, we compared performance across a spectrum of synchronization setting, which we term ``asynchronicity modes'' (Table \ref{tab:asynchronicity_modes}).
Asynchronicity mode 0 represents traditional fully-synchronous methodology.
Under this treatment, full barrier synchronization was performed between each computational update.
Asynchronicity mode 3 represents fully asynchronous methodology.
Under this treatment, individual threads or processes performed computational updates freely, incorporating input from other threads or processes on a fully best-effort basis.  

During early development of the library, we discovered that unprocessed messages building up faster than they could be processed --- even if they were being skipped over to only get the latest message --- could degrade quality of service or even cause runtime instability.
We opted for MPI communication primitives that could consume many backlogged messages per call and increased buffer size to address these issues, but remained interested in the possibility of partial synchronization to clear potential message backlogs.
So, we included two partially-synchronized treatments: asynchronicity modes 1 and 2.

In asynchronicity mode 1, threads and processes alternated between performing computational updates for a fixed-time duration and executing a global barrier synchronization.
For the graph coloring benchmark, work was performed in 10ms chunks.
For the digital evolution benchmark, which is more computationally intensive, work was performed in 100ms chunks.
In asynchronicity mode 2, threads and processes executed global barrier synchronizations at predetermined time points.
In both experiments, global barrier synchronization occurred on each time a second of epoch time elapsed.

Finally, asynchronicity mode 4 disables all inter-thread and inter-process communication, including barrier synchronization.
We included this mode to isolate the impact on performance of communication between threads and processes from other factors potentially affecting performance, such as cache crowding.
In this run mode for the graph coloring benchmark, all calls to put messages into or pull messages out of ducts between processes or threads were skipped (except after the benchmark concluded, when assessing solution quality).
Because of its larger footprint, incorporating logic into the digital evolution simulation to disable all inter-thread and inter-process messaging was impractical.
Instead, we launched multiple instances of the simulation as fully-independent processes and measured performance of each.

\subsection{Code, Data, and Reproducibility}

Benchmarking experiments were performed on [redacted for double-blind review]'s High Performance Computing Center, a cluster of hundreds of heterogeneous x86 nodes linked with InfiniBand interconnects.
For multithread experiments, benchmarks for each thread count were collected from the same node.
For multiprocess experiments, each processes was assigned to a distinct node in order to ensure results were representative of performance in a distributed context.
All multiprocess benchmarks recorded from the same collection of nodes.
Hostnames are recorded for each benchmark data point.
For an exact accounting of hardware architectures used, these hostnames can be crossreferenced with a table included with the data that summarizes the cluster's node configurations.

Code for the distributed graph coloring benchmark is available at \url{https://github.com/mmore500/conduit} under \\ \texttt{demos/channel\_selection}.
Code for the digital evolution simulation benchmark is available at \url{https://github.com/mmore500/dishtiny}.
Exact versions of software used are recorded with each benchmark data point.
Data is available via the Open Science Framework at \url{https://osf.io/7jkgp/} \cite{foster2017open}.
An in-browser notebook for all reported statistics and data visualizations and is available via Binder at \url{https://mybinder.org/v2/gh/mmore500/conduit/HEAD?filepath=binder\%2F} \cite{jupyter2018binder}.

\section{Results and Discussion}

\subsection{Multithread Benchmarks}

  \begin{figure}[thpb]
      \centering
    \begin{subfigure}[b]{\linewidth}
      \centering
      \includegraphics[width=\linewidth]{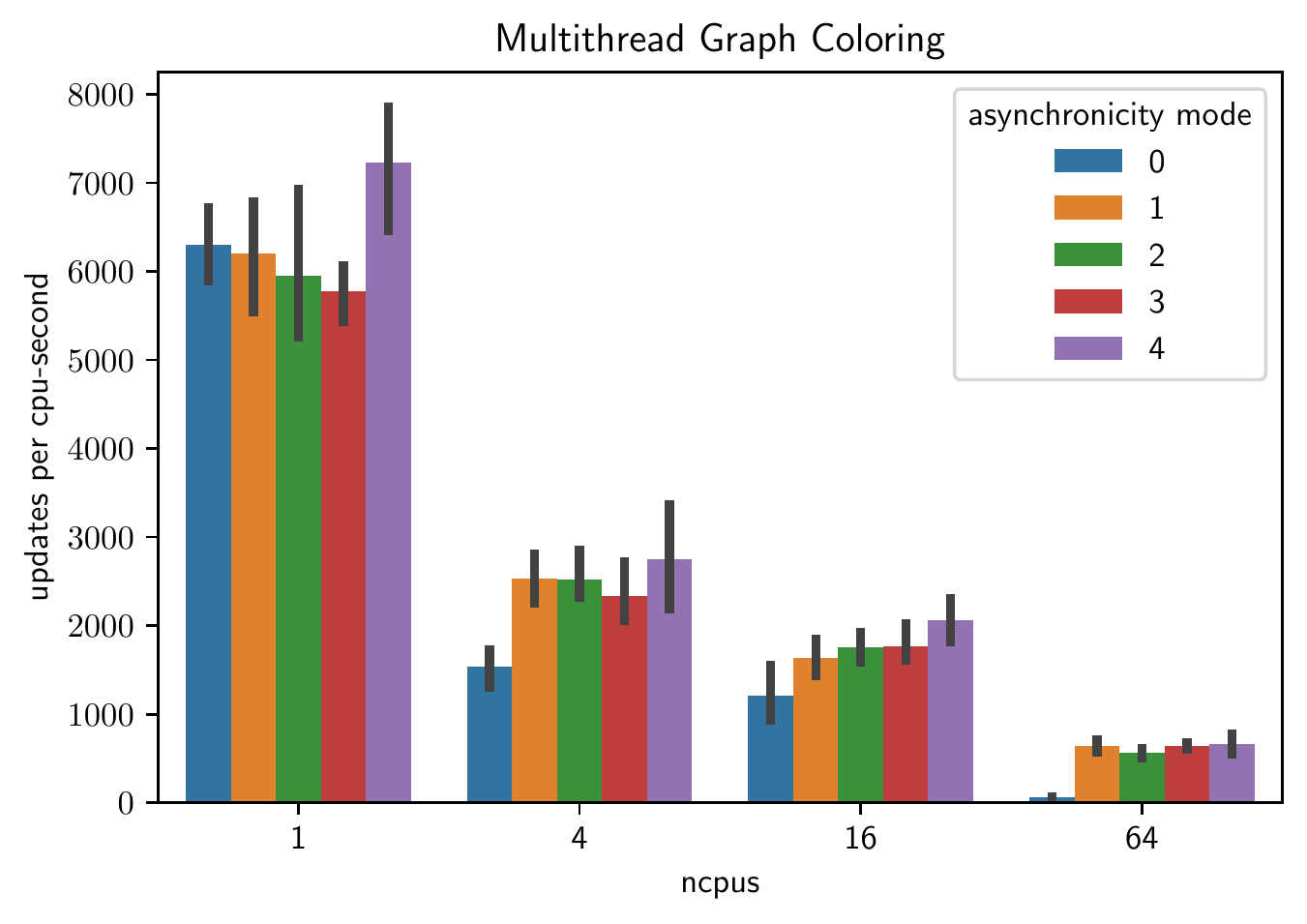}
     \caption{Graph coloring per-thread update rate. Higher is better.}
         \label{fig:multithread_graph_coloring_update_rate}
      \end{subfigure}
      
    \begin{subfigure}[b]{\linewidth}
      \centering
      \includegraphics[width=\linewidth]{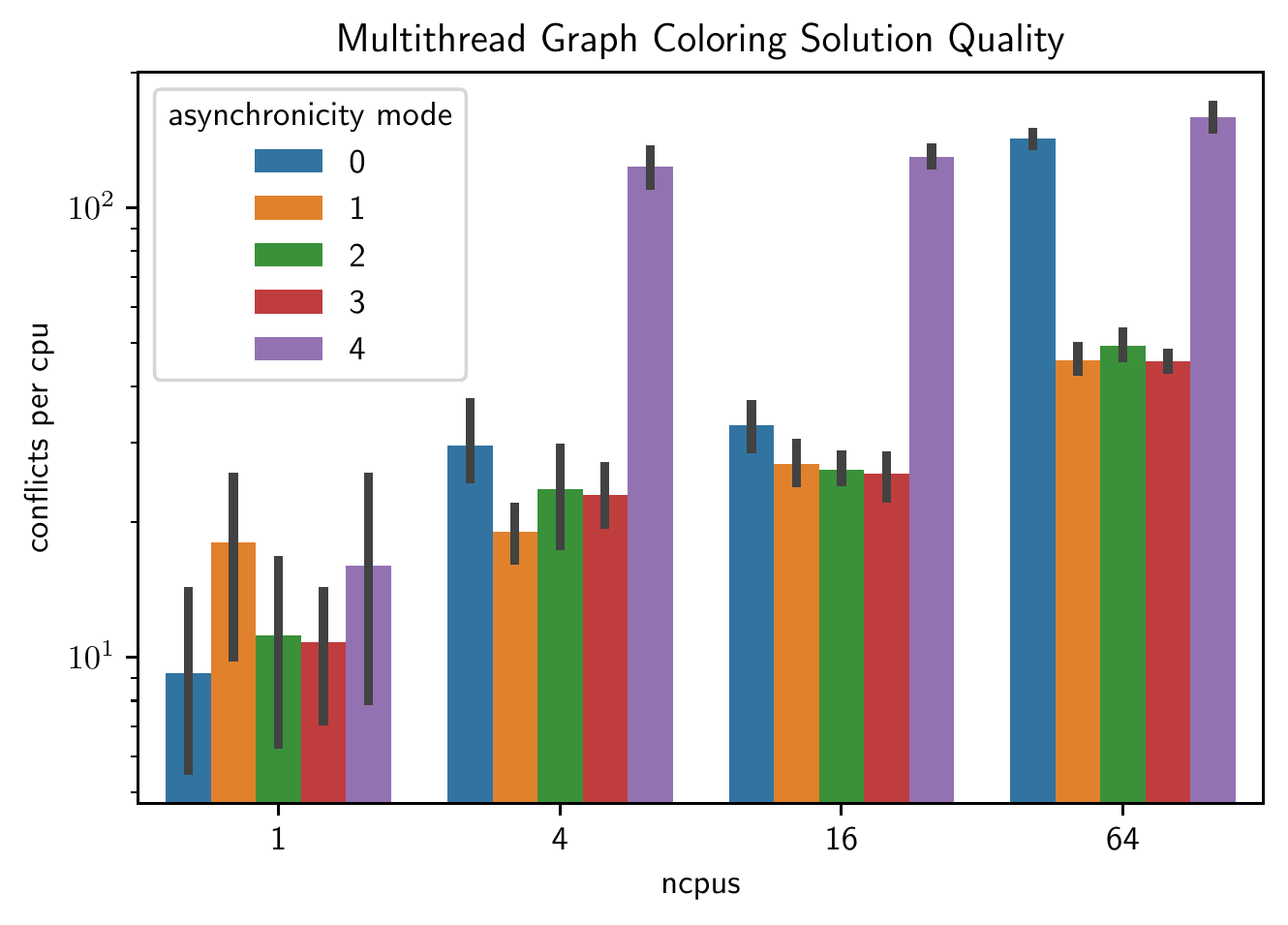}      
      \caption{Graph coloring solution conflicts. Lower is better.}
         \label{fig:multithread_graph_coloring_solution_quality}
    \end{subfigure}

    \begin{subfigure}[b]{\linewidth}
    \includegraphics[width=\linewidth]{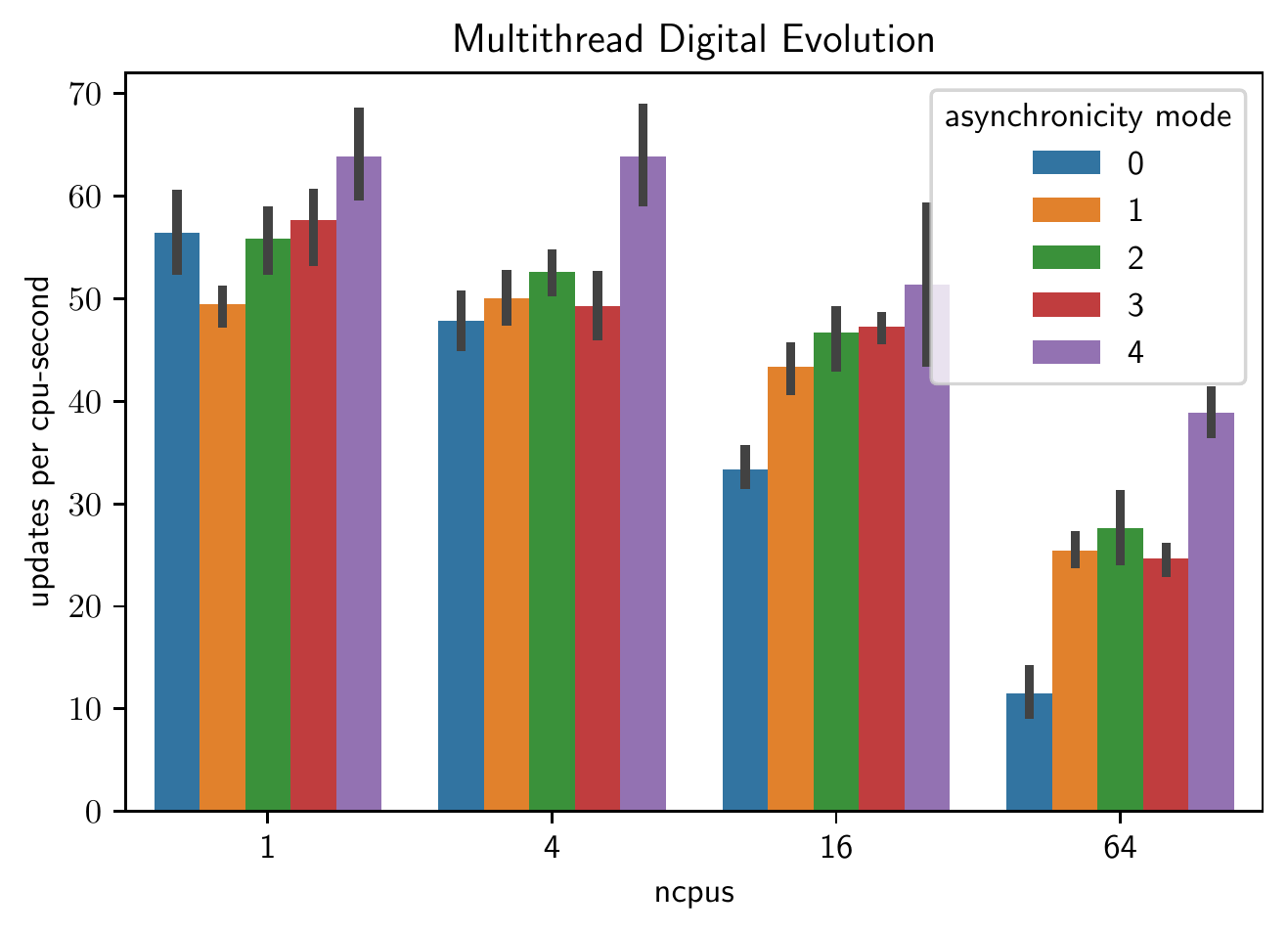}
    \caption{Digital evolution per-thread update rate. Higher is better.}
         \label{fig:multithread_digital_evolution_update_rate}
    \end{subfigure}

    \caption{Multithread benchmark results. Bars represent bootstrapped 95\% confidence intervals. }
      \label{fig:multithread_benchmarks}
  \end{figure}

Figure \ref{fig:multithread_graph_coloring_update_rate} presents per-cpu algorithm update rate for the graph coloring benchmark at 1, 4, 16, and 64 threads.
Update rate performance decreased with increasing multithreading across all asynchronicity modes.
This performance degradation was rather severe --- per-cpu update rate decreased by 61\% between 1 and 4 threads and by about another 75\% between 4 and 64 threads.
Surprisingly, this issue appears largely unrelated to inter-thread communication, as it was also observed in asynchronicity mode 4, where all interthread communication is disabled.
Perhaps per-cpu update rate degradation under threading was induced by strain on a limited system resource like memory cache or access to the system clock (which was used to control run timing).
This unexpectedly severe phenomenon merits further investigation to fully in future work with this benchmark.

Nevertheless, we were able to observe significantly better performance of best-effort asynchronicity modes 1, 2, and 3 at high thread counts. 
At 64 threads, these run modes significantly outperformed the fully-synchronized mode 0 ($p < 0.05$, non-overlapping 95\% confidence intervals).
Likewise, as shown in Figure \ref{fig:multithread_graph_coloring_solution_quality}, best-effort asynchronicity modes were able to deliver significantly better graph coloring solutions within the allotted compute time than the fully-synchronized mode 0 ($p < 0.05$, non-overlapping 95\% confidence intervals).

Figure \ref{fig:multithread_digital_evolution_update_rate} shows per-cpu algorithm update rate for the digital evolution benchmark at 1, 4, 16, and 64 threads.
Similarly to the graph coloring benchmark, update rate performance decreased with increasing multithreading across all asynchronicity modes --- including mode 4, which eschews inter-thread communication.
Even without communication between threads, with 64 threads each thread performed updates at only 61\% the rate of a lone thread.
At 64 threads, best-effort asynchronicity modes 1, 2, and 3 exhibit about 43\% the update-rate performance of a lone thread.
Although best-effort inter-thread communication only exhibits half the update-rate performance of completely decoupled execution at 64 threads, this update-rate performance is roughly $2.1\times$ that of the fully-synchronous mode 0.
Indeed, best-effort modes significantly outperform the fully-synchronous mode on the digital evolution benchmark at both 16 and 64 threads ($p < 0.05$, non-overlapping 95\% confidence intervals).

\subsection{Multiprocess Benchmarks}

\begin{figure}[thpb]
      \centering
      
    \begin{subfigure}[b]{\linewidth}
    \centering
    \includegraphics[width=\linewidth]{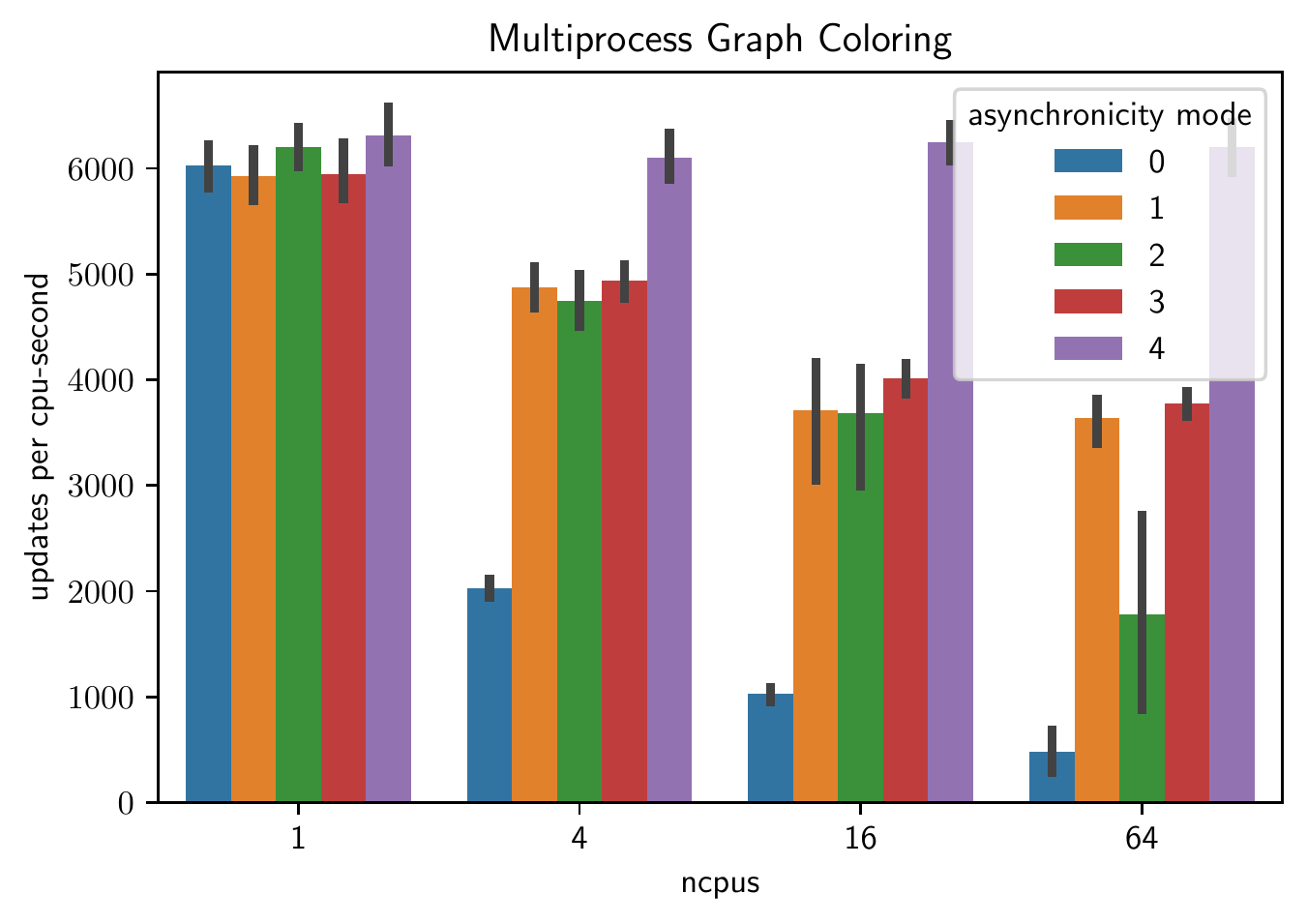}
    \caption{Graph coloring per-process update rate. Higher is better.}
    \label{fig:multiprocess_graph_coloring_update_rate}
    \end{subfigure}
    
    \begin{subfigure}[b]{\linewidth}
      \centering
      \includegraphics[width=\linewidth]{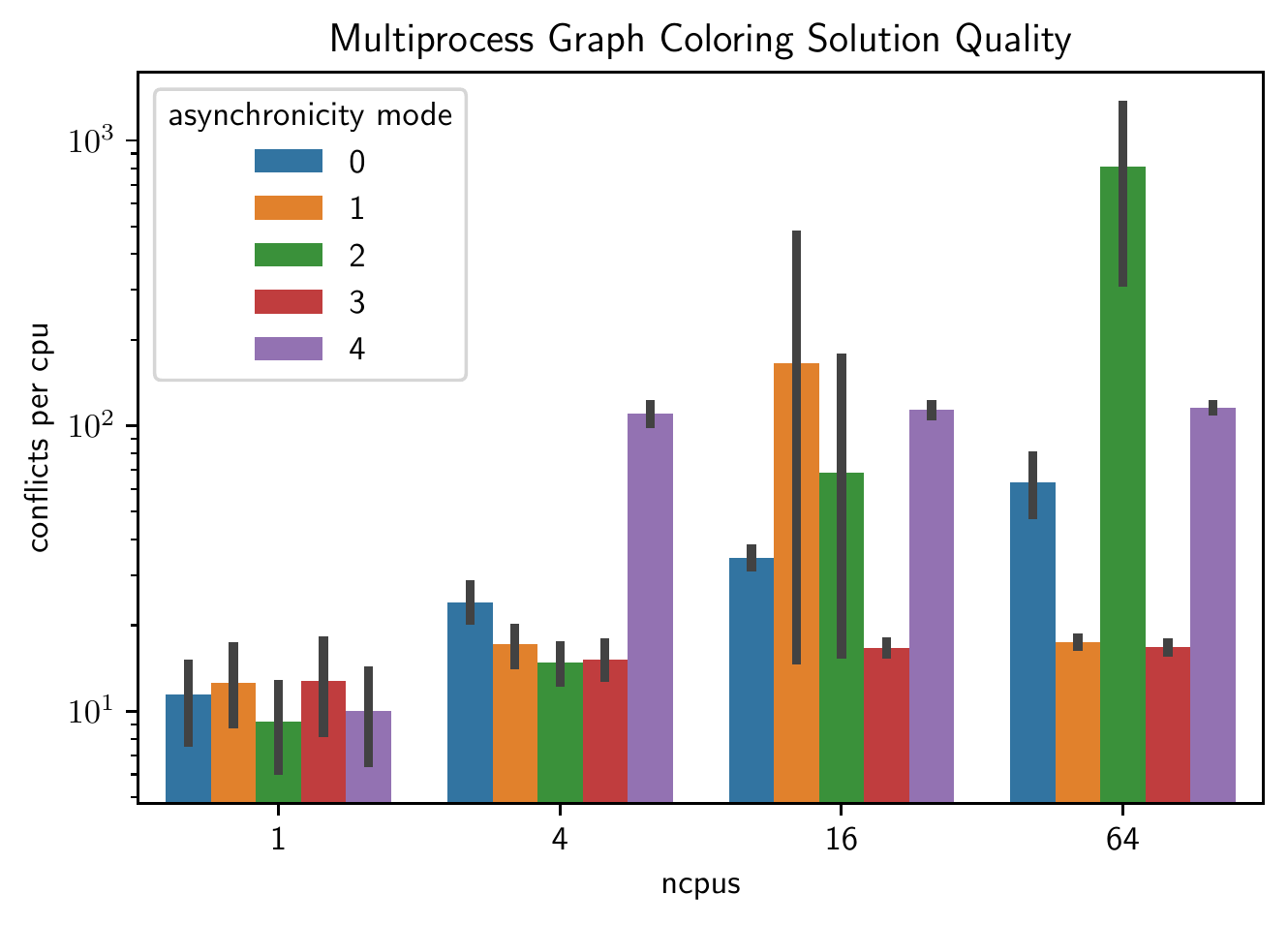} 
      \caption{Graph coloring solution conflicts. Lower is better.}
      \label{fig:multiprocess_graph_coloring_solution_quality}
    \end{subfigure}

  \begin{subfigure}[b]{\linewidth}
    \centering
  \includegraphics[width=\linewidth]{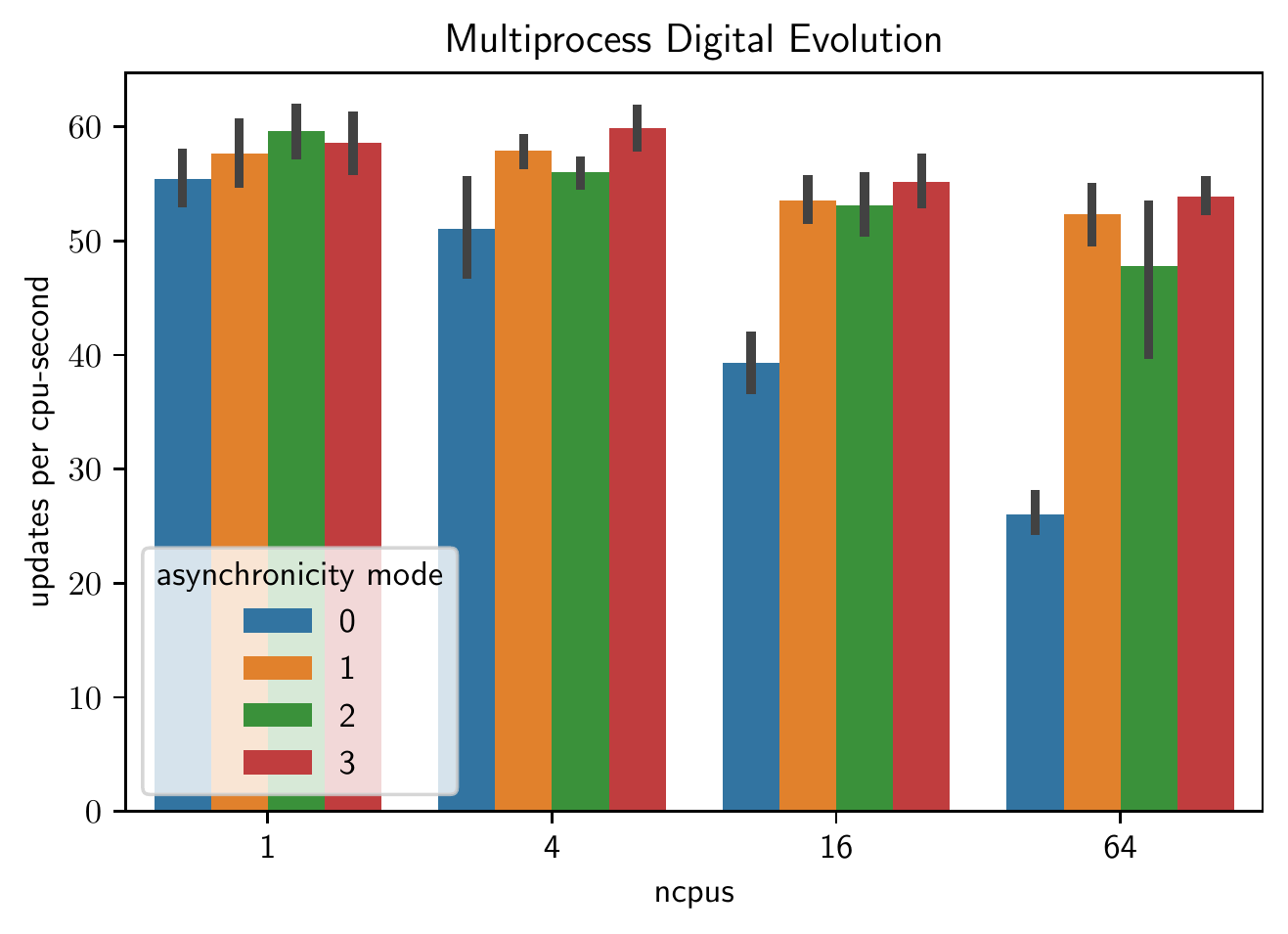}
  \caption{Digital evolution per-process update rate. Higher is better.}
  \label{fig:multiprocess_digital_evolution_update_rate}
  \end{subfigure}
      
  \caption{Multiprocess benchmark results. Bars represent bootstrapped 95\% confidence intervals. }
  \label{fig:multiprocess_benchmarks}
\end{figure}

Figure \ref{fig:multiprocess_graph_coloring_update_rate} shows per-cpu algorithm update rate for the graph coloring benchmark at 1, 4, 16, and 64 processes.
Unlike the multithreaded benchmark, multiprocess graph coloring exhibits consistent update-rate performance across process counts under asynchronicity mode 4, where inter-thread communication is disabled.
This matches the expectation that, indeed, with comparable hardware a single process should exhibit the same mean performance as any number of completely decoupled processes.
At 64 processes, best-effort asynchronicity mode 3 exhibits about 63\% the update-rate performance of single-process execution. 
This represents about an $7.8\times$ speedup compared to fully-synchronous mode 0.
Indeed, best-effort mode 3 enables significantly better per-cpu update rates at 4, 16, and 64 processes ($p < 0.05$, non-overlapping 95\% confidence intervals).

Likewise, shown in Figure \ref{fig:multiprocess_graph_coloring_solution_quality}, best-effort asynchronicity mode 3 yields significantly better graph-coloring results within the allotted time at 4, 16, and 64 processes ($p < 0.05$, non-overlapping 95\% confidence intervals).
Interestingly, partial-synchronization modes 1 and 2 exhibited highly inconsistent solution quality performance at 16 and 64 process count benchmarks.
Fixed-timepoint barrier sync (mode 2) had particularly poor performance performance at 64 processes (note the log-scale axis).
We suspect this was caused by a race condition where workers would assign sync points to different fixed points different based on slightly different startup times (i.e., process 0 syncs at seconds 0, 1, 2... while process 1 syncs at seconds 1, 2, 3..).

Figure \ref{fig:multiprocess_digital_evolution_update_rate} presents per-cpu algorithm update rate for the digital evolution benchmark at 1, 4, 16, and 64 processes.
Relative performance fares well at high process counts under this relatively computation-heavy workload, with 64-process fully best-effort simulation exhibiting about 92\% the update rate of single-process simulation.
This represents a $2.1\times$ speedup compared to the fully-synchronous run mode 0.
Best-effort significantly mode 3 outperforms the per-cpu update rate of fully-synchronous mode 0 at process counts 16 and 64 ($p < 0.05$, non-overlapping 95\% confidence intervals).

\section{Conclusion}

Benchmarks show that best-effort communication through Conduit enables significantly better computational performance under high thread and process counts.
We also demonstrated how, in the case of the graph coloring benchmark, best-effort communication can help achieve tangibly better solution quality within a fixed time constraint, as well.  
We observed the greatest relative speedup under distributed communciation-heavy workloads --- about $7.8\times$ on the graph coloring benchmark.
Distributing the computation-heavy digital evolution benchmark workload across independent nodes yielded the strongest scaling of our benchmarks, achieving at 64 processes 92\% the update-rate of single-process execution. 

In future work, we plan to further characterize the performance of the Conduit's best-effort model with respect to the digital evolution simulation, looking directly at quality of service metrics such as message latency and frequency of dropped messages.
We are also eager to investigate how Conduit's best-effort communication model scales on much larger process counts, perhaps on the order of hundreds or thousands of cores.

Development of the Conduit library stemmed from a practical need for an abstract, prepackaged interface to support our digital evolution research.
We hope that making this library available to the community can reduce domain expertise and programmability barriers to taking advantage of the best-effort communication model to efficiently leverage burgeoning parallel and distributed computing power. 

\begin{acks}
\footnotesize
This research was supported in part by NSF grants DEB-1655715 and DBI-0939454 as well as by Michigan State University through the computational resources provided by the Institute for Cyber-Enabled Research.
This material is based upon work supported by the National Science Foundation Graduate Research Fellowship under Grant No. DGE-1424871.
Any opinions, findings, and conclusions or recommendations expressed in this material are those of the author(s) and do not necessarily reflect the views of the National Science Foundation.

\end{acks}

\bibliographystyle{ACM-Reference-Format}
\bibliography{main}

\end{document}